\documentclass[reprint,aps,pre,floatfix]{revtex4-1}

\usepackage{amsfonts}
\usepackage{amsmath}
\usepackage{amssymb}
\usepackage{graphicx}
\usepackage{times}
\usepackage{hyperref}
\usepackage{epsfig}
\usepackage{dcolumn}
\usepackage{epstopdf}
\usepackage{circuitikz}
\usepackage{tikz}

\newcommand{\bear}{\begin{eqnarray}}
\newcommand{\eear}{\end{eqnarray}}
\newcommand{\be}{\begin{equation}}
\newcommand{\ee}{\end{equation}}
\newcommand{\Eqref}[1]{Eq.~(\ref{#1})}

\newcommand{\ud}{\mathrm{d}}
\newcommand{\im}{\mathrm{i}}
\newcommand{\nn}{\nonumber}
\newcommand{\jm}{\mathrm{j}}
\newcommand{\kb}{k_\mathrm{_B}}

\begin{document}

\title{Callen-Welton fluctuation dissipation theorem and Nyquist theorem as a consequence of detailed balance principle applied to an oscillator}

\author{T.~M.~Mishonov}
\email[E-mail: ]{mishonov@gmail.com}
\affiliation{Faculty of Physics,\\
St.~Clement of Ohrid University at Sofia,\\
5 James Bourchier Blvd., BG-1164 Sofia, Bulgaria}

\author{I.~M.~Dimitrova}
\affiliation{Faculty of Chemical Technologies, University of Chemical Technology and Metallurgy,\\
8 Kliment Ohridski blvd, BG-1756 Sofia}

\author{A.~M.~Varonov}
\email[E-mail: ]{akofmg@gmail.com}
\affiliation{Department of Theoretical Physics, Faculty of Physics,\\
St.~Clement of Ohrid University at Sofia,\\
5 James Bourchier Blvd., BG-1164 Sofia, Bulgaria}

\date{November 9, 2018}

\begin{abstract}

We re-derive the Nyquist theorem and Callen-Welton fluctuation-dissipation theorem (FDT) 
as a consequence of detailed balance principle applied to a harmonic oscillator.
The usage of electrical notions in the beginning makes the consideration understandable for every physicists. 
Perhaps it is the simplest derivation of these well-known theorems in statistical physics.  
The classical limit is understandable as a consequence of Waterston-Herapath 
equipartition theorem.
\end{abstract}

\maketitle

\section{Introduction}
\label{Sec:Introduction}
For half a century the general Callen-Welton\cite{Callen:51} fluctuation dissipation theorem (FDT) has been included in every professionally written course in statistical physics and is an indispensable part of the physics education.
See for example Landau-Lifshitz course on theoretical physics~\cite{LL5,LL9,LL10}.
Roughly speaking, FDT is a consequence of thermal averaging of the rate of quantum mechanical transitions from perturbation theory. 

On the other hand, the Nyquist theorem~\cite{Nyquist:28} for the thermal noise of the resistors is in the basis of the theory of electronic devices and even one can read in the literature that Nyquist theorem is an interesting application of the fluctuation dissipation theorem~\cite{LL9}. 
Actually the opposite is also true.
The purpose of this paper is demonstrate that FDT can be derived as a consequence of Nyquist theorem
applying the principle of detailed balance to a high-quality electric resonator.
And then to observe that there is nothing specifically electrical in this construction.
Our article is oriented to the reader which is not a specialist in FDT and quantum statistics in general.
Our purpose is to demonstrate a simple but correct re-derivation of  FDT which is understandable for the broader auditorium of physicists which use results of FDT in their current research or they teach it to application oriented students.
In any case we present a new derivation of the well-known theorem.
In the next section we will recall the basic notions related to fluctuations.
Introducing standard notations for spectral density of electric variables Sec.~\ref{Sec:Densities}
and considering an oscillator in electric variables Sec.~\ref{Sec:LRC}
we re-derive Nyquist theorem in Sec.~\ref{Sec:Nyquist}
and Collen-Welton theorem Sec.~\ref{FDT} 
in general case after some change of the notations in Sec.~\ref{Sec:Last change}.
Finally we arrive at the conclusion that our sequential derivation of those well-known from the textbooks theorems is perhaps the simplest one understandable for every BSc physicist.

\section{Spectral densities of the electric variables}
\label{Sec:Densities}
The model with which we start to think is the electric noise of a resistor.
Let for simplicity suppose that voltage noise $U(t)$ is a periodic function with long enough period
$U(t+\mathcal{T})=U(t)$ and we can use Fourier representation
\begin{align}
&U(t)= \sum_\omega U_\omega\exp(-\mathrm{i}\omega t), 
\qquad \omega=\frac{2\pi}{\mathcal{T}}n,\\
&U_\omega=\left<\exp(\mathrm{i}\omega t)U(t)\right>
=\frac1{\mathcal{T}}
\int_0^\mathcal{T}\exp(\mathrm{i}\omega t^\prime) U(t^\prime)\mathrm{d}t^\prime,
\end{align}
where $n$ is integer $n=0,\pm1, \pm2, \dots$.
One can introduce also the mean value 
\begin{align}
\langle U \rangle=\frac1{\mathcal{T}}\int_0^\mathcal{T} U(t^\prime)\mathrm{d}t^\prime
\end{align}
and dispersion
\begin{align}
(\delta U)^2=\langle (U- \langle U \rangle)^2 \rangle
=\frac1{\mathcal{T}}\int_0^\mathcal{T} (U- \langle U \rangle)^2
\mathrm{d}t^\prime.
\end{align}
For thermal noise $\langle U \rangle=0$ and analogously the mean current is also zero
$\langle I \rangle=0.$
When statistic properties of the noise are constant one can introduce spectral density $(U^2)_f$
as the integrand of the frequency $f=\omega/2\pi$ integration
\begin{align}
(\delta U)^2=\langle (U- \langle U \rangle)^2 \rangle
=\int_0^\infty (U^2)_f\frac{\mathrm{d}\omega}{2\pi}.
\end{align}
The subscript $f$ in the spectral density means that integration is with respect to the Hz frequency $f=\omega/2\pi$. In the Landau-Lifshitz course of theoretical physics~\cite{LL5} the used subscript is $\omega$, but in our work we follow the accepted in the electronics application and engineering folding of negative and positive frequencies $\omega$ in the even spectral density, i.e. $(U^2)_f=2(U^2)_\omega.$ 
One can consider that subscript $f$ means also folding.
This minor difference in the notations does not influence the final result -- FDT.
For the averaged energy of a capacitor with capacitance $C$ we have
\begin{align}
\left<E_C\right>=\frac{C}2\left<U^2\right>.
\end{align}
In the next section we will consider the relations between spectral densities of a
high quality electric resonator.

\section{Conductivity of high-Q sequential LRC-resonance circuit}
\label{Sec:LRC}

\begin{figure}[h]
\includegraphics[scale=0.4]{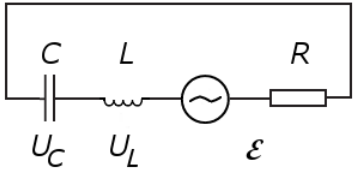}
\caption{Resonance circuit with inductance $Z_L=\jm \omega L$, capacitor $Z_C=1/\jm \omega C$ and resistor $Z_R=R$ creating random electric voltage with spectral density given by Nyquist theorem
$(\mathcal{E}^2)_f=2 \hbar \omega \kb T^\prime \coth(\hbar \omega/2 \kb T^\prime)$, where $\kb$ is the Boltzmann constant.}
\label{fig:osc}
\end{figure}

In Fig.~\ref{fig:osc} a resonance LC circuit with a resistor $R$ creating random noise is depicted.
The random thermal voltage $\mathcal{E}_\omega$ creates a current amplitude
$I_\omega=\sigma(\omega) \mathcal{E}(\omega)$, where the conductivity 
$\sigma(\omega)=1/Z(\omega)$ is determined by the total impedance $Z=Z_{LRC}$ 
of the sequentially connected inductance $L$, resistor $R$ and capacitor $C$
\be
Z_{LRC}(\omega)=\jm \omega L + R + \frac{1}{\jm \omega C},
\quad \mathrm{j}=-\mathrm{i}.
\label{RLC_impedance}
\ee
Let us recall that in the general case for arbitrary frequency dependence
the Ohmic resistance is the real part of the complex impedance $R(\omega)=\Re (Z(\omega))$.
As a gedanken experiment, let us analyze a high quality resonance circuit, for which
\be
R \ll \sqrt{\omega L\frac{1}{\omega C}}=\sqrt{\frac{L}{C}},
\quad \mathcal{Q}\equiv\frac{\sqrt{L/C}}{R}\gg 1.
\ee
In this case the square of the modulus of the conductivity
\be
\left| \sigma(\omega) \right|^2=\frac{1}{R^2+\left(\omega L-\frac{1}{\omega C}\right)^2} \approx
\frac{\pi}{2} \frac{1}{RL} \delta(\omega-\omega_0)
\label{sig2}
\ee
has a sharp maximum at the resonance frequency $\omega_0 = 1/\sqrt{LC}$ and it is negligible far from the resonance. In such a way we can use the $\delta$-function approximation 
\be
F(\omega_0)=\int_0^\infty F(\omega)\delta(\omega-\omega_0)\mathrm{d}\omega,
\quad \omega_0>0.
\ee
The coefficient in front of the delta function in Eq.~(\ref{sig2}) is given by the integral
\be
\int_0^\infty \left| \sigma(\omega) \right|^2 \mathrm{d}\omega
=\int_0^\infty \frac{\omega^2 \ud \omega}{R^2 \omega^2 + (L \omega^2 -1/C)^2}=
\frac{\pi}{2} \frac{1}{RL},
\label{iintegral}
\ee
which does not depend on the capacitance $C$.
Introducing a dimensionless variable $x \equiv \omega L/R$
and a dimensionless parameter $\mathcal{Q}^2=L/CR^2$,
the corresponding mathematical problem is reduced to the calculation of the integral
\be
\int_{-\infty}^\infty \frac{x^2 \ud x}{x^2+(x^2-\mathcal{Q}^2)^2} = \pi,
\label{jntegral}
\ee
which was done analytically a century ago~\cite{Schottky:18,Schottky:22}.
The use of electric variables is not indispensable, oscillator can be analyzed in mechanical variables as well~\cite{LL1}.

\section{Relations between spectral densities of high quality resonator. Nyquist theorem}
\label{Sec:Nyquist}

Using for $\mathcal{Q}\gg1$ the $\delta$-function approximation 
for the square of modulus of the conductivity 
Eq.~(\ref{sig2}) we obtain for the spectral density of the current
\be
(I^2)_f =\vert\sigma(\omega)\vert^2(\mathcal{E}^2)_f
\approx \frac{\pi}{2} \frac{(\mathcal{E}^2)_f}{RL} \delta(\omega-\omega_0).
\label{CurrentDensity}
\ee
The general relation between spectral densities is irrelevant from the mechanics, quantum or classical.
Now we can calculate the average energy of the inductance
\be
\left < E_L\right > = \frac{L}{2} (\delta I)^2, \quad (\delta I)^2 = \int_0^\infty (I^2)_f \frac{\mathrm{d} \omega}{2 \pi}, 
\label{CurrenDispersion}
\ee
and the integration of the $\delta$-function gives
\be
\left < E_L\right >= \frac{L}{2} \cdot \frac{\pi}{2RL} (\mathcal{E}^2)_f \cdot \frac1{2\pi},
\label{eq:EL}
\ee
where to alleviate re-derivation central dots separate factors from different origin.
On the other hand, according to the virial theorem for the oscillator the averaged energy of the inductance is equal to the averaged energy of the capacitor $\left < E_C \right >=\left < E_L\right >$ and their sum must be equal to the averaged energy of a quantum oscillator
\begin{align}
&\left < E_C \right > + \left < E_L\right > = \bar{\varepsilon}, \label{eq:vir} \\
&\bar{\varepsilon}=\frac12 \hbar \omega_0 \coth \left (\frac{\hbar \omega_0}{2T} \right )=
\frac{\sum\limits_{n=0}^\infty E_n \mathrm{e}^{-E_n/T}}{\sum\limits_{n=0}^\infty \mathrm{e}^{-E_n/T}}, \\
&E_n= \hbar \omega_0 \left ( n + \frac12 \right ), \quad T=\kb T^\prime,
\end{align}
where $\kb$ is the Boltzmann constant.
The discrete levels of the quantum oscillator mean that $Q$ is actually 
the quantum operator of the electric charge of the capacitor $\hat{Q}$.

For high temperatures we have the equipatition theorem
\be
\bar{\varepsilon} \approx \kb T^\prime \gg \hbar \omega_0,\qquad
\left < E_L \right >\approx\left < E_C \right >\approx \frac12 k_\mathrm{B}T^\prime.
\ee

The 
substitution of $\left < E_L \right >$ from \Eqref{eq:EL} in \Eqref{eq:vir} gives the Nyquist theorem
\be
(\mathcal{E}^2)_f = 4 R \bar{\varepsilon},
\label{eq:Nyq}
\ee
which have the well-known form for the  the Johnson noise at high temperatures
gives the well-known formula for
\be
(\mathcal{E}^2)_f = 4 R \kb T^\prime.
\label{Nyquist}
\ee
In such a way we re-derived this difficult Nyquist theorem applied for the classical case
$h f\ll T.$
For the important classical case applicable practically for all electronic circuits, 
the spectral density of the electric noise $(\mathcal{E}^2)_f = 4 R \kb T^\prime$
can be proved as the only function of frequency giving equiparticipation theorem for arbitrary 
$\mathcal{Q}$.
The nature of the conductivity is irrelevant for this theorem we have ona and the same spectral density of the electric noise for metals with quantum transport of electrons an electrolytes with classical motion of ions.

In the general case we have derived the Nyquist theorem analyzing the detailed balance~\cite{LL10} of a high-Q resonator instead of the original Nyquist~\cite{Nyquist:28} consideration of one dimensional black body radiation in a transmission line and effectively one dimensional black body radiation.
Substituting the mean energy of a quantum oscillator in the Nyquist theorem 
for the spectral density of the thermal voltage of the resistor reads
\be
(\mathcal{E}^2)_f = 2 R(\omega) \hbar \omega \coth \left ( \frac{\hbar \omega}{2 \kb T^\prime} \right ),
\ee
where by definition the Ohmic resistance $R(\omega)=\Re(Z(\omega))$ 
is the real part of the complex impedance for every circuit.
Rederiving Nyquist theorem we have to emphasize the conditions. 
One and the same frequency dependent resistance $R(\omega)$ (the real part of the impedance by definition)
is used in both the quantum $\kb T^\prime\ll \hbar \omega$
and classical cases $\hbar \omega\ll \kb T^\prime$ for different temperatures $T^\prime$ 
at a fixed frequency $\omega$; 
this statement has never been disputed, 
see Ref.~\onlinecite{Nyquist:28,Callen:51,LL5}.
The mechanics of charge carriers is as a rule quantum, imagine electrons moving in metal. 
But the particular frequency dependence $Z(\omega)$ is irrelevant for the general theorem.
The general theorem gives a relation between two independently measured variables:
the spectral density of the voltage noise $(\mathcal{E}^2)_f $
and the real part of the frequency dependent impedance $R(\omega)$.
One and the same impedance is used in both quantum and classical case 
and that is why it is better to trace the thermal averaging of the perturbation theory
related to the Callen-Welton theorem~\cite{Callen:51,LL5}. 
Implicitly Nyquist\cite{Nyquist:28} made the same, ignoring the 
nature of the impedance and its real part $R(\omega)$,
he derived the general relation using the detailed balance principle applied for a transmission line.
We have rederived the same result using not a transmission line but a resonance circuit.
Again, the nature of the impedance $Z(\omega)$, quantum or classical, is irrelevant.

Now as a test of our approach we will check the equipartition theorem.

\subsection{Waterston-Herapath equipartition theorem~\cite{Idiot}}
For the derivation of the Nyquist theorem we use $\delta$-function approximation for the high-Q resonator.
In this section we derive the energies of the capacitor and inductance for the general case
and will verify that Nyquist formula for the spectral density at high temperatures is compatible with the equipartition theorem.

For the classical statistics $\hbar\omega\ll \kb T$ substitution the spectral density of the voltage noise
from Eq.~(\ref{Nyquist}) into the formula for the spectral density of the current 
Eq.~(\ref{CurrentDensity}) $(I^2)_f =\vert\sigma(\omega)\vert^2(\mathcal{E}^2)_f$ and later into the formula for the current dispersion Eq.~(\ref{CurrenDispersion}) using the integral 
Eq.~(\ref{iintegral}) gives the classical equpartition theorem for the mean energy of the thermal currents
flowing through an inductance
\be
\frac12 L\left<I^2\right>=\frac12 \kb T^\prime.
\ee  

\subsection{Average energy of the capacitor}
The averaging of the energy of the capacitor
\be
\left < E_C \right > = \frac{\left < Q^2 \right >}{2C}, \quad 
\left < Q^2 \right > = \int_0^\infty (Q^2)_f \, \mathrm{d}f
\ee
is analogous.
The spectral density of the electric charge 
$(Q^2)_f = (I^2)_f/\omega^2$
can be expressed by the spectral density of the current.
The substitution in the formula for the energy 
\be
\left < E_C \right > = \frac{1}{2C} \int_0^\infty  \frac{(I^2)_f}{\omega^2} \frac{\mathrm{d} \omega }{2 \pi}
\ee
taking into account the formula for the current density \ref{CurrentDensity} 
and Nyquist result for the voltage noise density~\ref{Nyquist} give
\be
\left < E_C \right > = \frac{1}{2C} 4 R \kb T^\prime \int_0^\infty \frac{\vert \sigma(\omega) \vert^2} {\omega^2} \frac{\mathrm{d} \omega }{2 \pi}.
\ee
Here we can use the formula for the conductivity Eq.~(\ref{sig2}) and arrive at the integral
\begin{align}
& \left < E_C \right > = \frac{4 R T}{2C} \frac{L}{R^3} \frac{J}{4\pi}, \label{ec} \\
&J \equiv 2 \int_0^\infty \frac{\ud \omega}{R^2 \omega^2 + (L \omega^2 -1/C)^2}.
\end{align}
Introducing the dimensionless frequency $x=\omega L/R$ and and square of the quality factor
$\mathcal{Q}^2=L/CR^2$
using the integral by Schottky~\cite{Schottky:22,Schottky:18}
\be
J = \int_{-\infty}^\infty \frac{\ud x}{x^2 + (x^2 -\mathcal{Q}^2)^2}=\frac{\pi}{\mathcal{Q}^2}.
\label{j}
\ee
and substituting \Eqref{j} in \Eqref{ec} and the expression for $\mathcal{Q}$
we finally arrive at the equiparitition theorem for the energy of the capacitor
\be
\left < E_C \right > = \frac12 \kb T^\prime.
\ee

In the next section we will represent the impedance by the capacitance.

\section{Expressing the impedance by the capacitance. FDT}
\label{FDT}
For every linear circuit containing not only passive elements $R$, $L$ and $C$
but also transistors and operational amplifiers, the impedance is defined as a ratio
between the small complex amplitudes of the current and voltage 
\begin{align}
Z(\omega)=\frac{U_0}{I_0},\quad U(t)=U_0\mathrm{e}^{\mathrm{j}\omega t}
,\quad I(t)=I_0\mathrm{e}^{\mathrm{j}\omega t}, \quad |I_0|\rightarrow 0.
\end{align}
Quantum or classical, the nature of the processes creating the impedance is irrelevant for our further consideration.
Having complex functions from real frequency $\omega$ every impedance
can be represented by frequency dependent capacitance or inductance
\be
Z(\omega)=\frac{\im}{\omega C(\omega)}=-\mathrm{i}\omega L(\omega).
\label{GeneralImpedance}
\ee
While in Sec.~\ref{Sec:LRC}  $L$ and $C$ are supposed to be ideal capacitance and inductance 
of a resonance circuit here $L(\omega)$ and $C(\omega)$ are different representations the frequency dependent inductance $Z(\omega)$ in general case i.e.
here we are obliged to repeat that the influence of all capacitors, resistors and impedances has not been neglected but has been included in the common impedance of the whole circuit, which can be represented by the frequency dependent capacitance or inductance. 
As a rule, the choice of the notions depends on the low frequency behavior
of the circuit impedance. 
However, in our case in order to derive the FDT, we need to arrive to a Hamiltonian and 
that is why we have to represent the impedance by the frequency dependent capacitance.
In some sense it is only a change of the notations.
The resistance also can be represented by the capacitance
\be
\omega R = \omega \Re\left ( \frac{\im}{\omega C} \right )=\frac{C^{\prime\prime}}{|C|^2}, \quad
C=C^\prime+\im C^{\prime\prime},
\ee
and the Nyquist theorem can be rewritten as
\be
(\mathcal{E}^2)_f 
=2 \frac{\hbar C^{\prime\prime}}{|C|^2} \coth \left ( \frac{\hbar \omega}{2 \kb T^\prime} \right ).
\label{eq:E2f}
\ee
On the other hand the spectral density of the electric charge can be expressed by the frequency dependent capacitance and the spectral density of the voltage 
$(Q^2)_f = |C(\omega)|^2 (\mathcal{E}^2)_f$ 
and here substitution of $(\mathcal{E}^2)_f$ from \Eqref{eq:E2f} gives
\be
(Q^2)_f = 2\hbar C^{\prime\prime}(\omega) \coth \left ( \frac{\hbar \omega}{2 \kb T^\prime} \right ).
\ee
In such a way for the dispersion of the charge for zero thermal averaged value $\left < Q \right >=0$ 
we finally obtain 
\begin{align}
&\left < Q^2 \right > = (\delta Q)^2 = \int_0^\infty (Q^2)_f \frac{\mathrm{d}\omega}{2 \pi} \nn \\
&=\frac{\hbar}{\pi} \int_0^\infty C^{\prime\prime} (\omega) \coth \left ( \frac{\hbar \omega}{2 \kb T^\prime} \right ) \mathrm{d} \omega.
\label{eq:NyQ}
\end{align}

The relation between the charge,
actually the time dependent averaged value of the 
charge operator $\overline{Q}(t)\equiv\left<\hat{Q}\right>$
and the external voltage $U(t)$ this is a classical variable which creates the deviation from the 
thermal equilibrium value 
in general case can be written in time-representation
\be
\overline{Q}(t)=\int_0^\infty \tilde{C}(\tau)U(t-\tau) \ud \tau
=\int_{-\infty}^t \tilde{C}(t-t^\prime)U(t^\prime) \ud t^\prime
\label{eq:Qt}
\ee
and frequency-representation
\be
Q_\omega=C(\omega)U_\omega,
\ee
where 
\be
\tilde{C}(t)=\int_{-\infty}^{+\infty} \mathrm{e}^{-\im \omega t} C(\omega) \frac{\ud \omega}{2 \pi}, \quad
C(\omega)=\int_{-\infty}^{+\infty} \mathrm{e}^{\im \omega t} \tilde{C}(t) \ud t.
\ee
We suppose that at zero perturbation $U(t)=0$ the thermal averaged mean value of the charge operator is zero
$\left<\hat{Q}\right>=0$. 
We use one and the same letter for different functions which represent one and the 
the same physical variable in different representations; time dependent $\tilde{C}(t)$ and 
frequency dependent $C(\omega)$.
These functions are connected with the reversible Fourier transformations.
But in the context of the problem, it is obvious that $C(\omega)$ 
from Eq.~\ref{GeneralImpedance} is the same frequency dependent capacitance for every circuit.
In frequency representation $C(\omega)$ has no singular point in the upper complex semi-plane
and in time representation $\tilde{C}(\tau)=\theta(\tau)\tilde{C}(\tau).$ 
This is the general principle of causality, applied for every electronic circuit and every physical system:
the response cannot precede the cause.
The time dependent kernel of the generalized capacitance $C(t)$ represents the response of the circuit;
for the trivial case of Ohm law we have, for example
\begin{align}
Q(t)=\int_{-\infty}^t \frac{U(t^\prime)}{R} \ud t^\prime,
\quad \tilde{C}_R(t-t^{\prime})=\frac{1}{R}.
\end{align}
While for an ideal capacitor we have a $\delta$-function like kernel
\begin{align}
&Q(t)=C U(t)=\int_{-\infty}^t  C \delta(t-t^{\prime}) \ud t^\prime,\\
&\tilde{C}_\mathrm{ideal}(t-t^{\prime})=C\delta(t-t^{\prime}).\nn
\end{align}
The $\delta$ in the formula above is as a rule omitted in the textbooks. 
We give special examples for the general kernel of the capacitance $\tilde{C}(t)$.
In Eq.~(\ref{RLC_impedance}) the quantities $R$, $L$, and $C$ represent the parameters of ideal
resistor, inductance and capacitor and $Z_{LRC}$ is the impedance of a sequentially connected 
resonance circuit.
While in Eq.~(\ref{GeneralImpedance} it is shown how the
frequency dependent impedance $Z(\omega)$ of every circuit can be represented by the
frequency dependent capacitance $C(\omega)$ or inductance $L(\omega)$.
Let us distinguish the special example from the general notions.

Imagine now that the voltage $\mathcal{E}=Q_0/C_0$ applied to the capacitor is taken from a large capacitor (charge reservoir) with initial charge $Q_0$.
The energy of the whole system is
\begin{align}
&E=\frac{Q^2}{2C}+\frac{(Q_0-Q)^2}{2C_0}=\frac{Q^2}{2C}-\mathcal{E}Q + \frac{Q_0^2}{2C_0}+\frac{Q^2}{2C_0} \nn \\
&=\left(\frac1{C}+\frac1{C_0}\right)\frac{Q^2}{2} - \mathcal{E}Q + \frac{Q_0^2}{2C_0},\\
& \approx \frac{Q^2}{2C} - \mathcal{E}Q +\mathrm{const}.
\quad C \ll C_0.
\end{align}
We can interpret this as a capacitor with energy $\hat{H}^{(0)}=Q^2/2C$ perturbed by an external voltage 
and the energy of this perturbation is
\be
V=-\mathcal{E}(t) Q,
\label{eq:V}
\ee
cf. [\onlinecite{LL5}]. 
The term $Q^2/2C_0$ is negligible because we consider that the capacitance $C_0\rightarrow \infty$ and $1/C_0 \ll 1/C$, 
i.e. we treat $C_0$ as a charge reservoir or source of a voltage $Q_0/C_0$.
The last term $\propto Q_0^2$ is a irrelevant for our consideration constant.
In our general analysis we do not use any special property of the impedance for which we derived the Nyquist theorem, that is why these results are applicable for all impedances but actually the for all one dimensional systems. Let us make the final change of the notations.

\section{Change of the notations}
\label{Sec:Last change}

In order to emphasize the generality of the derived results, we can change the notations.
We can denote the capacity by $\alpha$, 
the charge by the indifferent letter $x$ appropriate to be used for an arbitrary degree of freedom, 
and the electromotive force can be denoted by $F(t)$ which is typical notation for force in mechanics.
Performing this change of the notations
\begin{align}
Q \rightarrow x,\quad C(\omega)\rightarrow \alpha(\omega), 
\quad U\rightarrow F
\end{align}
the Nyquist theorem \Eqref{eq:NyQ} can be rewritten as
\be
\left < x^2 \right >=\frac{\hbar}{\pi} \int_0^\infty \alpha^{\prime\prime} (\omega) 
\coth \left ( \frac{\hbar \omega}{2 \kb T^\prime} \right ) \mathrm{d} \omega,
\ee
which is exactly the Callen--Welton fluctuation-dissipation theorem (FDT)
in the notation of \cite{LL5}.
For the response we have analogously from \Eqref{eq:Qt}
\be
\overline{x}(t)=\int_0^\infty \alpha(\tau) F(t-\tau) \ud \tau
\ee
and this result is applicable for every linearly perturbed system 
with the perturbation Hamiltonian and \Eqref{eq:V} reads as
\be
\hat{V}=-\hat{x} F(t);
\ee
now is clear why we have represented the impedance by the capacitance.
As we have already pointed out, above the interval of time integration $\tau>0$ 
is the general consequence of the causality principle: the response of the mean system, i.e. value 
averaged operator $\overline{x}$ at moment $t$
depends from the values of the external force $F$ from the preceding (not following) time moments.

\section{Discussion an conclusion}
\label{Sec:Discussion}

When the Nyquist theorem is derived from the Hamiltonian 
approach using standard Gibbs averaging of the perturbation theory, 
the Nyquist theorem takes the final name Callen-Welton fluctuation-dissipation theorem (FDT). 
And for such point of view the Nyquist theorem is \textit{an interesting application of FDT}~\cite{LL9}.

In such a way, we have made a closed loop in the notions 
and ideas related to the electric fluctuations 
and all other fluctuations. 
In our approach one oscillator is used, in the Nyquist approach a transmission line, 
while in the general Callen-Welton proof of the Nyquist theorem there are no model 
assumptions, thermodynamics considerations or principle of the detailed balance,
but state of the art methods of quantum and statistical mechanics. 
We hope that our approach also completes this chain 
and gives some methodological advantages for university education.
The derivations of the important for the statistical physics 
Nyquist and Callen-Welton theorems are now understandable 
for physicists having no expertise in FDT in quantum systems.

In conclusion, we re-derive both the Nyquist~\cite{Nyquist:28} and the Callen-Welton~\cite{Callen:51} theorems as a consequence of the detailed balance principle~\cite{LL10} without any additional assumptions.
The electrical notations used in the beginning are not indispensable,
the detailed balance principle can be applied using mechanical notations.
No recollection of the original literature is necessary;
we have cited the original articles and the well-known 
textbooks~\cite{LL5,LL9,McCombie:71,Kittel:80} and the future 
extended review by~\cite{Reggiani:19} 
only for comparison of length and technical details.
In such a way our derivation is understandable for every physicist having BSc degree.



\appendix

\end{document}